\documentclass[a4paper,11pt]{article}
\pdfoutput=1 

\usepackage{jheppub} 

\usepackage[T1]{fontenc} 

\title{\boldmath Conformality in many-flavour lattice QCD at strong coupling}

\author[a,b]{Ph. de Forcrand,}
\author[c]{S.~Kim}
\author[a,d]{and W.~Unger}

\affiliation[a]{Institut f\"ur Theoretische Physik, ETH Z\"urich, CH-8093 Z\"urich, Switzerland}
\affiliation[b]{CERN, Physics Department, TH Unit, CH-1211 Geneva 23, Switzerland}
\affiliation[c]{Department of Physics, Sejong University, Seoul 143-747, Korea}
\affiliation[d]{Institut f\"ur Theoretische Physik, Goethe-Universit\"at Frankfurt, 60438 Frankfurt am Main, Germany}

\emailAdd{forcrand@phys.ethz.ch}
\emailAdd{skim@sejong.ac.kr}
\emailAdd{unger@th.physik.uni-frankfurt.de}

\abstract{
It is widely believed that chiral symmetry is spontaneously broken at zero temperature in the strong coupling limit of staggered fermions, for any number of colors and flavors. 
Using Monte Carlo simulations, we show that this conventional wisdom, based on a mean-field analysis, is wrong. 
For sufficiently many fundamental flavors, chiral symmetry is restored via a bulk, first-order transition. 
This chirally symmetric phase appears to be analytically connected with the expected conformal window of many-flavor continuum QCD. 
We perform simulations in the chirally symmetric phase at zero quark mass 
for various system sizes $L$, 
and measure the torelon mass, the Dirac spectrum and the hadron spectrum. 
All masses go to zero with $1/L$. $L$ is hence the only infrared length scale.
Thus, the strong-coupling chirally restored phase appears as a convenient 
laboratory to study IR-conformality. Finally, we present a conjecture for the
phase diagram of lattice QCD as a function of the bare coupling and the number
of quark flavors.
}


\usepackage{bbm}
\usepackage{verbatim}
\usepackage{psfrag}
\usepackage{slashed}
\usepackage[normalem]{ulem}

\usepackage{amsmath}
\usepackage{amssymb}
\usepackage{marvosym}
\usepackage{stmaryrd}
\usepackage{wasysym}
\usepackage{mathcomp}
\usepackage{multirow}

\def \Nc {{N_{c}}}
\def \Nf {{N_{f}}}
\def \hatNf {\hat{N_{f}}}

\newcommand{\expval}[1]{\langle #1 \rangle}

\newcommand{\lr}[1]{\left( #1 \right)}

\newcommand{\beqn} {\begin{equation}}
\newcommand{\eqn} {\end{equation}}

\def \beq{\begin{equation}}
\def \eeq{\end{equation}}
\def \bea{\begin{eqnarray}}
\def \eea{\end{eqnarray}}

\def \Tr {{\rm Tr}}

\def \bet0{\beta_0}
\def \bet1{\beta_1}
\def \simgt{\,\rlap{\lower 7.5 pt\hbox{$\mathchar \sim$}}\raise 3 pt \hbox{$>$}\,}
\def \simlt{\,\rlap{\lower 7.5 pt\hbox{$\mathchar \sim$}}\raise 3 pt \hbox{$<$}\,}

\def\lsim{\raise0.3ex\hbox{$<$\kern-0.75em\raise-1.1ex\hbox{$\sim$}}}
\def\gsim{\raise0.3ex\hbox{$>$\kern-0.75em\raise-1.1ex\hbox{$\sim$}}}

\def \Ord{\mathcal{O}}

\def \pbp {\psi \bar{\psi}}

\begin{document}

\preprint{\texttt{\footnotesize CERN-PH-TH/2012-204}}
\maketitle

\section{Introduction}

The possibility that the Higgs boson could be a composite bound-state in
a high-energy Technicolor theory~\cite{TC} has generated considerable interest, especially
in the lattice community. In particular, the requirement that the Technicolor
theory be ``walking''~\cite{walk}, in order to accommodate stringent bounds on 
flavor-changing neutral currents, has been the driving motivation behind
several large-scale computer simulation efforts to determine the possible 
combinations of gauge groups and fermion contents leading to a conformal window.

To determine via lattice Monte Carlo simulations whether a given theory is 
inside the conformal window is
particularly challenging, because it involves a triple difficulty:
in order to identify (or not) an infrared fixed point (IRFP) which
is the signature of a theory inside the conformal window, one must probe 
the extreme infrared properties of the theory, while at the same time
taking the continuum limit of the lattice discretization, and controlling
the limit when the quarks become massless. This compounded difficulty
may explain why, in spite of considerable efforts, there is no consensus yet
on the minimum number $\Nf^*$ of quark flavors needed for QCD to be inside the
conformal window~\cite{review}.
A numerical demonstration of walking has been provided only recently,
in a toy model, the 2-$d$ $O(3)$ model at vacuum angle $\theta\approx\pi$~\cite{O3}.

Here, we relax the demand that results should be obtained in the continuum
limit. On a coarse lattice, long-distance properties can be studied more
economically. While such properties may differ from those of the 
corresponding continuum theory, it may still be instructive to consider
the possible existence of an IRFP for a discretized lattice theory.
The phase diagram of SU$(\Nc)$ gauge theory with $\Nf$ fundamental fermions,
as a function of $\Nf$ and the bare gauge coupling,
has been predicted in the celebrated Ref.~\cite{Miransky-Yamawaki},
which serves as a guide to understand the results of Monte Carlo simulations
performed at finite bare coupling. It is important to confront these
predictions with uncontroversial numerical evidence.
Therefore, we start our investigation by considering the strong coupling
limit, where the lattice is maximally coarse. 

Note that we consider standard staggered fermions, and (away from the 
strong-coupling limit) the standard plaquette action. Other discretizations
could lead to different results, since only the continuum limit is universal.

The conventional wisdom for strong coupling QCD with staggered fermions is that chiral symmetry remains always broken at zero temperature, 
regardless of the number of colors and flavors. This belief is based on 
mean-field analyses performed in some of the earliest papers on lattice QCD.
In particular, it was shown in \cite{Kluberg1983} that at leading order in 
a $1/d$ expansion, the chiral condensate has a value independent of the number
of colors $\Nc$ and of the number of staggered fields $\hatNf = \Nf/4$, where
$\Nf$ would be the corresponding number of degenerate fermion flavors in the
weak-coupling limit,
but depends only on the number $d$ of spatial dimensions:
\begin{align}
\expval{\pbp}(T=0) &=\sqrt{\frac{2}{d}}\left(1-\frac{1}{4d}\right)
\end{align}

Chiral symmetry may be restored by increasing the temperature $T$.
Following the approach of \cite{Damgaard1985}, where explicit results are
provided for a few small values of $\Nc$ and $\hatNf$, we calculated the chiral restoration temperature $aT_c$ and found that it is indeed non-zero for all $\hatNf$, and independent of $\Nc$ to leading order in $1/\hatNf$:
\begin{align}
aT_c&=\frac{d}{4}+\frac{d}{32}\frac{\Nc}{\hatNf}+\mathcal{O}\lr{\frac{1}{\hatNf^2}}
\end{align}
Hence chiral symmetry will never be restored at zero temperature, according to the mean-field analysis.
Since mean-field theory is expected to work well when the number of d.o.f. per site 
is large (e.g.~providing exact results in the Gross-Neveu model for $\Nf \rightarrow \infty$), 
there was no reason to doubt the validity of this finding.
Besides, it is in accord with the intuition that the gauge field is maximally
disordered in the strong-coupling limit, and that this disorder will drive
chiral symmetry breaking.

On the other hand, one naively should expect the above disorder to be modified 
by dynamical fermions, which have an ordering effect. 
Indeed, the loop expansion of the determinant shows that the fermionic 
effective action induced by dynamical fermions, 
$S_{\rm eff} = -\log\det(\slashed{D} + m_q)$, starts with a positive 
plaquette coupling $\Delta\beta$, proportional to $1/m_q^4$ for heavy quarks,
which has been studied numerically in 
\cite{Hasenfratz1994}. Clearly, for $\Nf$ flavors the effective action is
proportional to $\Nf$, and $\Delta\beta$ grows proportionally. 
This plaquette term suppresses fluctuations in the gauge field, 
which suggests that chiral symmetry restoration might take place for sufficiently large $\Nf$.

\section{Monte Carlo results}

\begin{figure}[htbp!]
\includegraphics[width=0.45\textwidth]{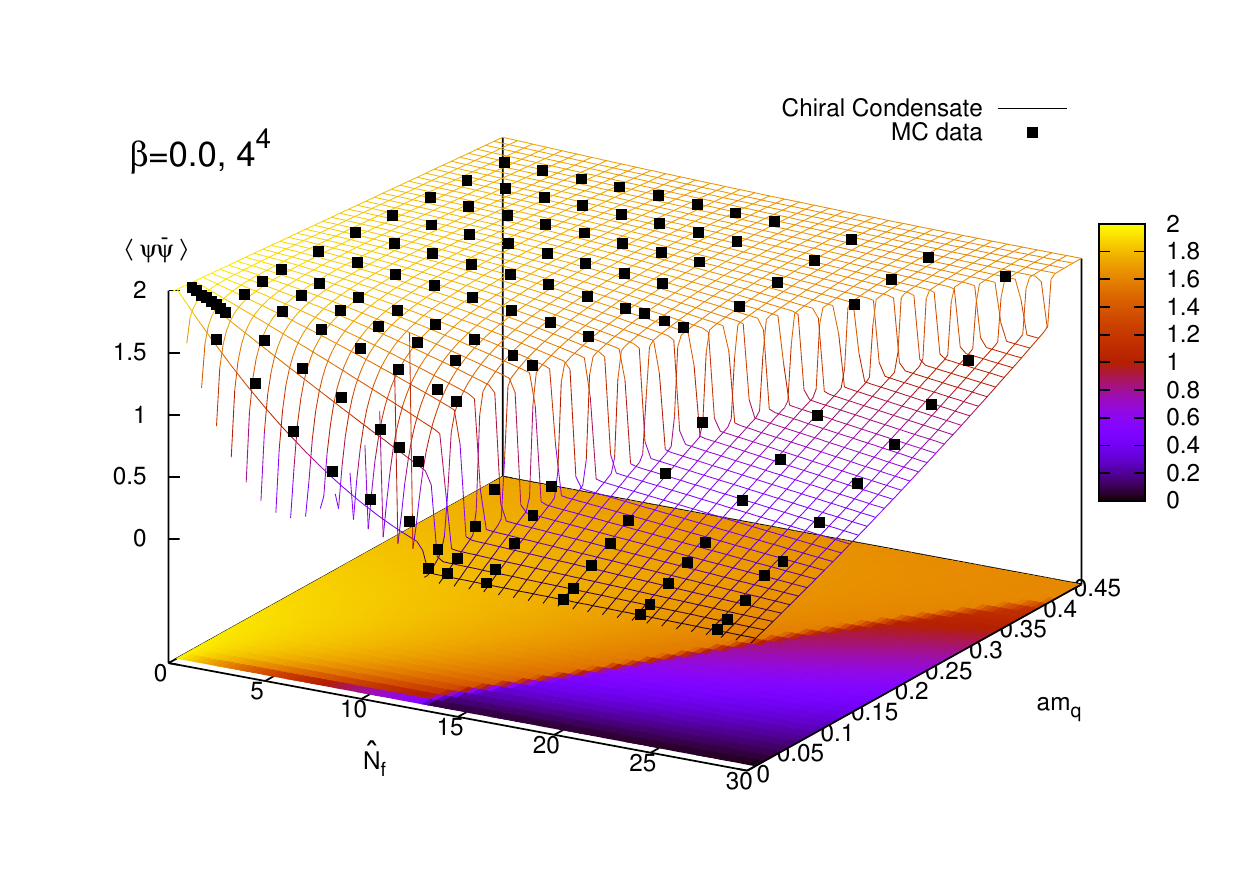}
\includegraphics[width=0.45\textwidth]{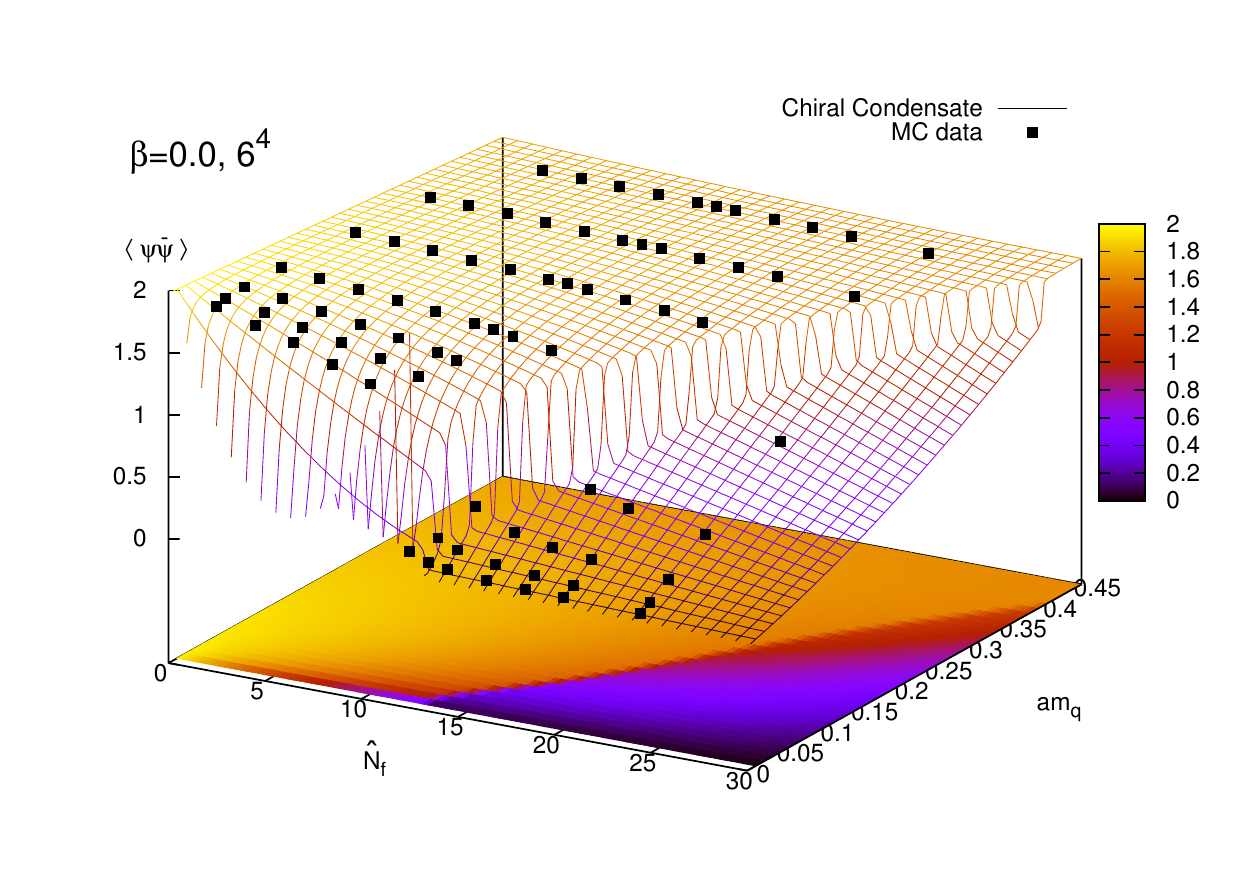}
\caption{The chiral condensate at strong coupling, $\beta=0$, in the $(\hatNf,am_q)$ plane, for $4^4$ (left) and $6^4$ (right) lattices.}
\label{beta0}
\includegraphics[width=0.45\textwidth]{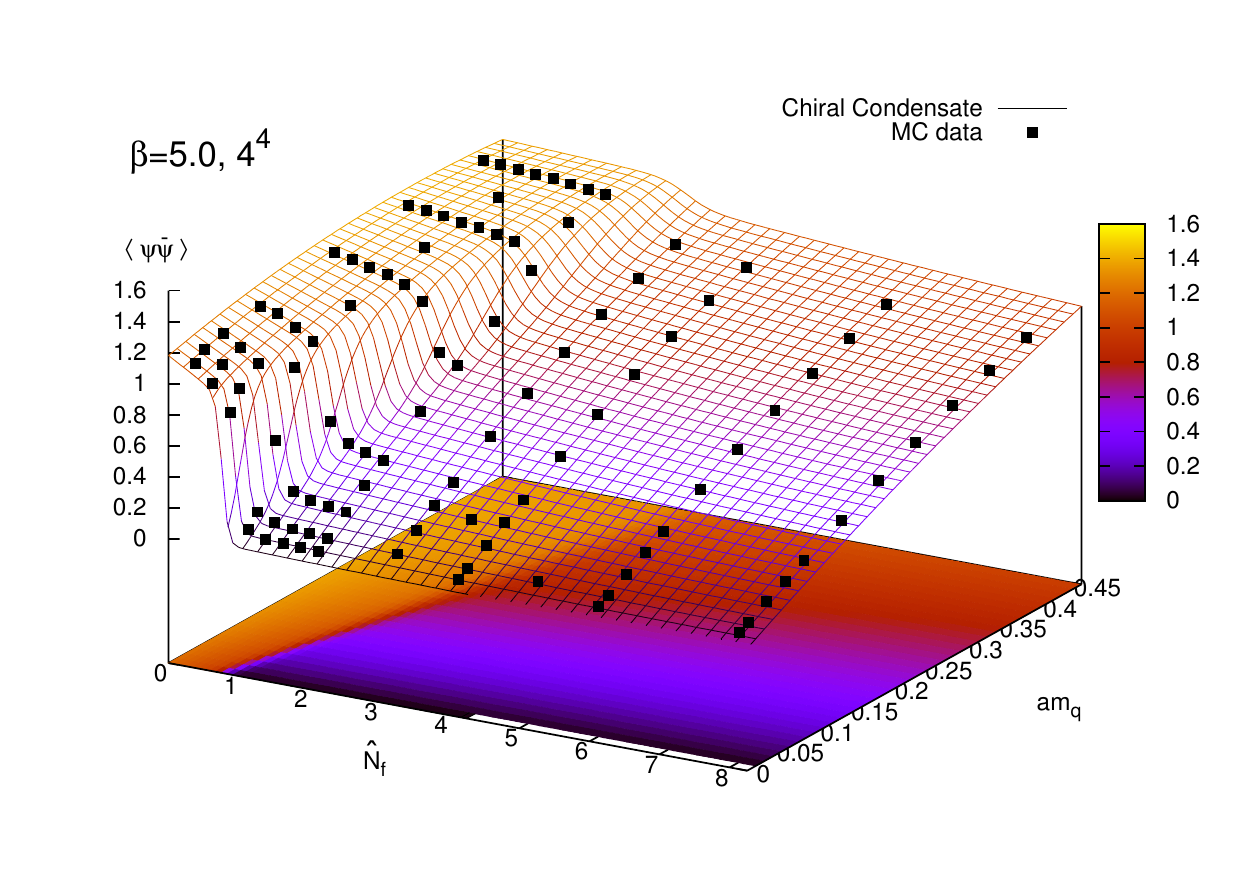}
\includegraphics[width=0.45\textwidth]{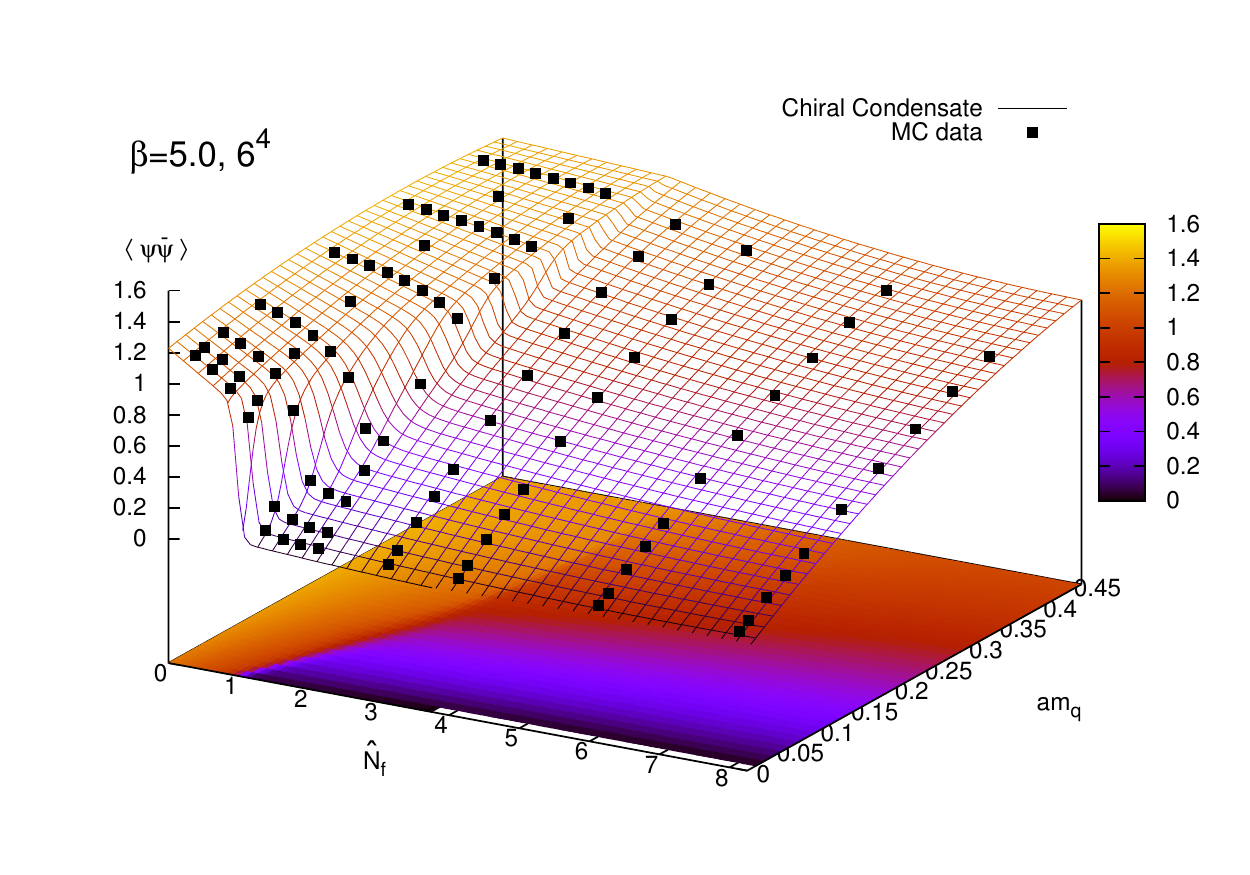}
\caption{The chiral condensate at weaker coupling, $\beta=5$, in the $(\hatNf,am_q)$ plane, for $4^4$ (left) and $6^4$ (right) lattices.}
\label{beta5}
\end{figure}

The only way to resolve this puzzle is to perform Monte Carlo simulations 
in the strong coupling limit of staggered fermions, to detect a possible chiral 
symmetry restoration for sufficiently large $\hatNf$. 
These simulations are straightforward, using the standard Hybrid Monte Carlo
algorithm.
As expected, the effect of increasing $\hatNf$ on the chiral condensate is to 
reduce its magnitude. But it came as a surprise to find that the chiral condensate vanishes 
via a strong first-order transition at $\hatNf^c\simeq 13$ staggered fields in the chiral limit (i.e. $\Nf^c \simeq 52$ continuum fermion flavors).
In the broken phase, the chiral condensate remains almost constant. It vanishes 
in the chiral limit due to finite-size effects only. 
In contrast, in the chirally restored phase the condensate is caused by
explicit symmetry breaking and is proportional to the quark mass.
This is illustrated Fig.~\ref{beta0}, where the condensate is shown
as a function of $\hatNf$ and bare quark mass $(a m_q)$.
Moreover, this $\Nf$-driven transition turns out to be a bulk, zero-temperature
transition, 
which can be seen by the fact that finite-size effects on the phase boundary are small when 
comparing two different system sizes $4^4$ and $6^4$, as shown Fig.~\ref{beta0} (left and right).

One also observes in these figures that the critical number of flavors increases
with the quark mass. This is easy to understand: heavier quarks have a weaker
ordering effect, so that the induced plaquette coupling $\Delta\beta$ 
decreases if one keeps $\hatNf$ fixed. It takes more flavors to keep the
system chirally symmetric. Hence,
$\hatNf^c$ increases, and for heavy quarks should obey $\hatNf^c\propto (a m_q)^4$.

One can now go back to the mean-field treatment and trace the origin of its
failure. Two kinds of terms at least are neglected: $(i)$ multiple meson
hopping along a given link, and $(ii)$ baryon loops. These terms amount
to corrections $\Ord(\hatNf/\Nc)$ and $\Ord(\hatNf/d^2)$, 
respectively, where $\hatNf=\Nf/4$ is the number of staggered fields, and
is normally set to 1 in the mean-field treatment. Here, we consider 
$\hatNf \gtrsim 13$, and the previously neglected corrections become
dominant. The conventional wisdom that chiral symmetry is always broken at $T=0$
in the strong-coupling limit comes from mistakenly applying the lowest-order
mean-field approximation in a regime where it is invalid.

Having established an $\hatNf$-driven phase transition in the strong-coupling
limit, we may consider its impact on the lattice theory at non-zero
lattice gauge coupling $\beta$ as well.
Since the transition is strongly first order, it has to persist for some range
in $\beta$ at least.
Hence we have compared the strong coupling phase diagram with the phase diagram at weaker coupling $\beta=5$, illustrated Fig.~\ref{beta5}. 
We find a similar qualitative behavior, but with
$\hatNf^c$ drastically reduced to $\Ord(2)$. Finite-size effects are more 
pronounced, but the transition still seems to be a first-order bulk transition.

\begin{figure}
\includegraphics[width=0.49\textwidth]{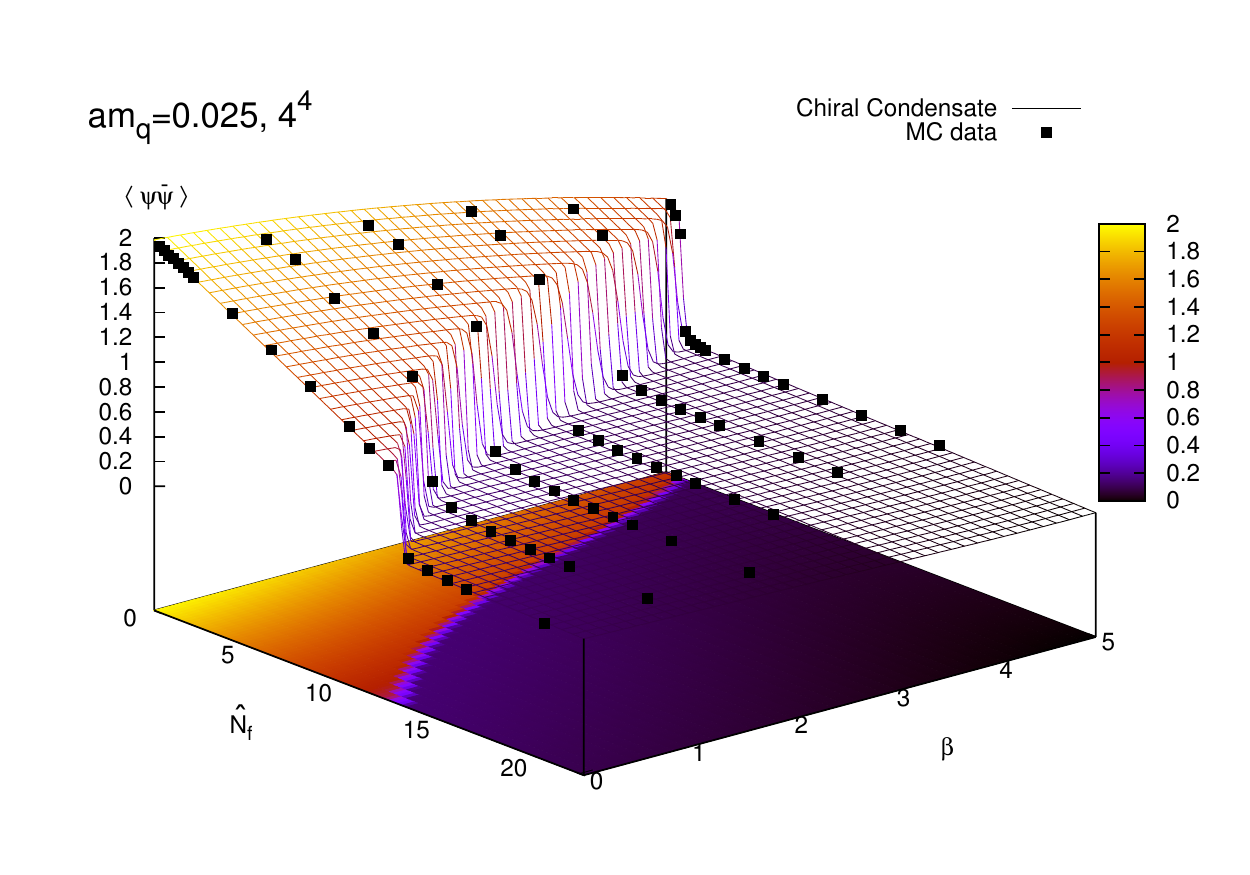}
\includegraphics[width=0.49\textwidth]{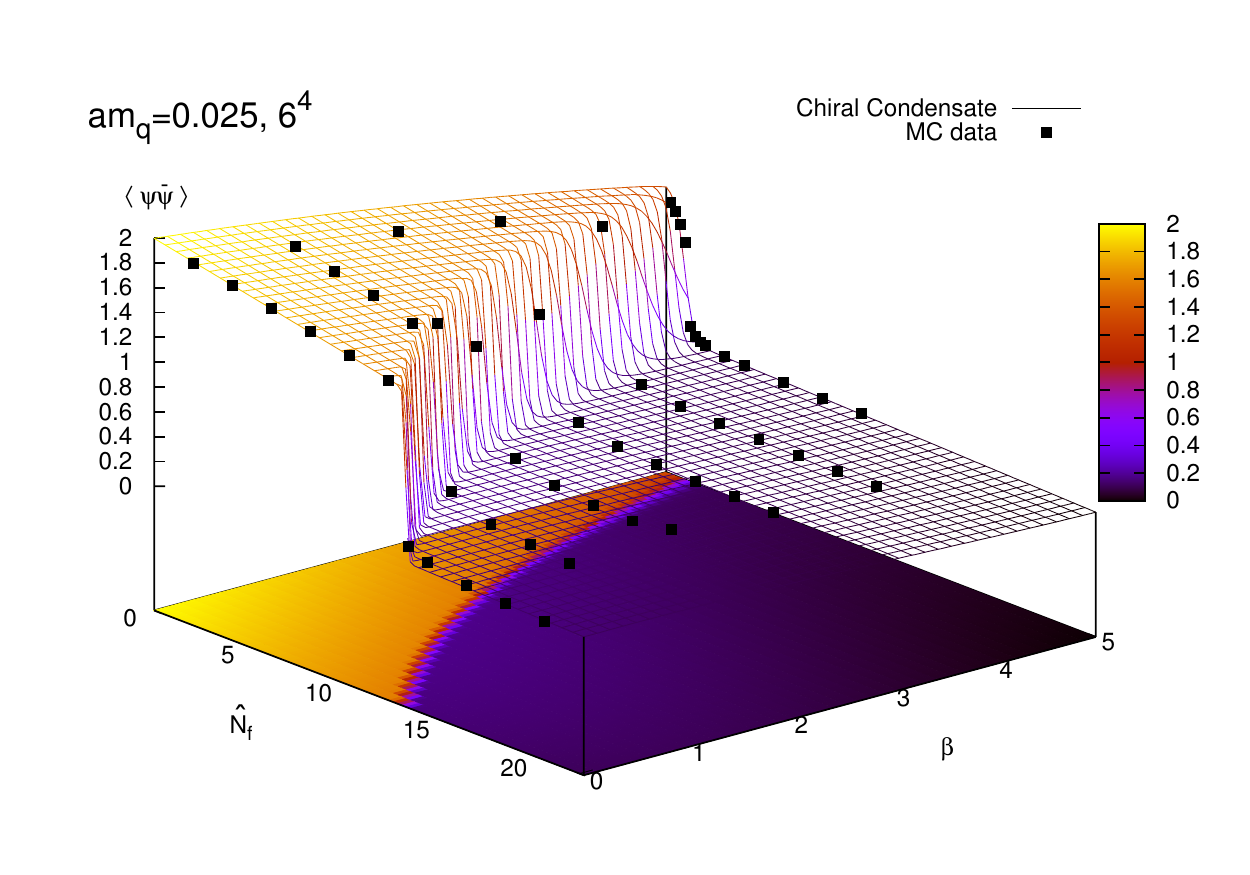}
\caption{The chiral condensate at $am_q=0.025$ in the $(\hatNf,\beta)$ plane, for 
$4^4$ (left) and $6^4$ (right) lattices. The phase transition remains strongly 
first order at weaker coupling.
The contour plot indicates the qualitative behaviour of the phase boundary, extending to weaker coupling.}
\label{fixedam}
\end{figure}

In fact, we find a smooth variation of the $\hatNf$-driven transition with $\beta$ at a given small quark mass $am_q=0.025$, as shown Fig.~\ref{fixedam}. 
The transition extends to weak coupling, at least to $\beta=5$, and remains 
strongly first-order. Thus, it is plausible that this transition, which
separates a chirally broken (small $\Nf$) and a chirally symmetric (large $\Nf$)
phase, persists all the way to the $\beta\to\infty$ continuum limit, where
it is to be identified with the transition at $\Nf=\Nf^*$ between the
chirally broken and the IR-conformal, chirally symmetric phase. 
In other words, our chirally restored phase may be analytically connected 
to the conformal window in the continuum limit, because
we do not observe any additional non-analyticity as $\beta$ is increased.

This possibility motivates our study of the properties of the strong-coupling
chirally symmetric phase, looking for tests of IR-conformality.

\section{Looking for conformality in the chirally symmetric phase}

\newcommand{\rank}{\text{rank}}

It is natural to ask whether the chirally restored phase is connected to the conformal window, i.e.~whether the chirally restored phase at strong coupling is also IR-conformal.
And if this is indeed the case, the next obvious question is whether this IR-conformal 
phase is trivial, ie.~whether the IR fixed point coupling is zero or not. 
In this section we present measurements of gluonic and fermionic observables 
chosen to address these
questions: the torelon mass from which we define a running coupling, 
the Dirac eigenvalue spectrum, and the hadron spectrum.
Our results support the following conclusion: the strong-coupling chirally 
symmetric phase is indeed IR-conformal, and it is non-trivial.

The simulations performed here are all in the chirally symmetric phase 
at zero plaquette coupling, with $\hatNf=14$ and $24$ staggered fields, which 
would correspond, in the weak-coupling limit, to
$\Nf=56$ and $96$ continuum flavors, 
and with lattices of size $4^3\times 16$,
$6^3\times 16$, $8^3\times 16$, $10^3\times 20$ and $12^3\times 24$.
The quark mass is set exactly to zero unless specified otherwise. We will see 
below that the Dirac operator has a spectral gap in the symmetric phase,
which makes a study of the massless theory within reach of modest computer
resources. Moreover, having one infrared scale, the system size $L$,
rather than two scales ($L$ and $1/m_q$) is of great advantage when analyzing
the results.

Let us mention the average plaquette values which we measure:
$\approx 0.35$ and $\approx 0.52$ for $\hatNf=14$ and $24$, respectively 
(normalized to 1 for the free field). So we are very far from a plaquette
value of 0, corresponding to maximally disordered gauge fields and achieved
for $\Nf=0$: the ordering effect of the dynamical fermions plays a dominant
role in our case, and the vanishing of the plaquette coupling is not
associated with special properties.

\subsection{Characterizing the chirally restored phase: (I) The Torelon Mass}

\begin{figure}[htbp!]
\centerline{
\includegraphics[width=0.45\textwidth]{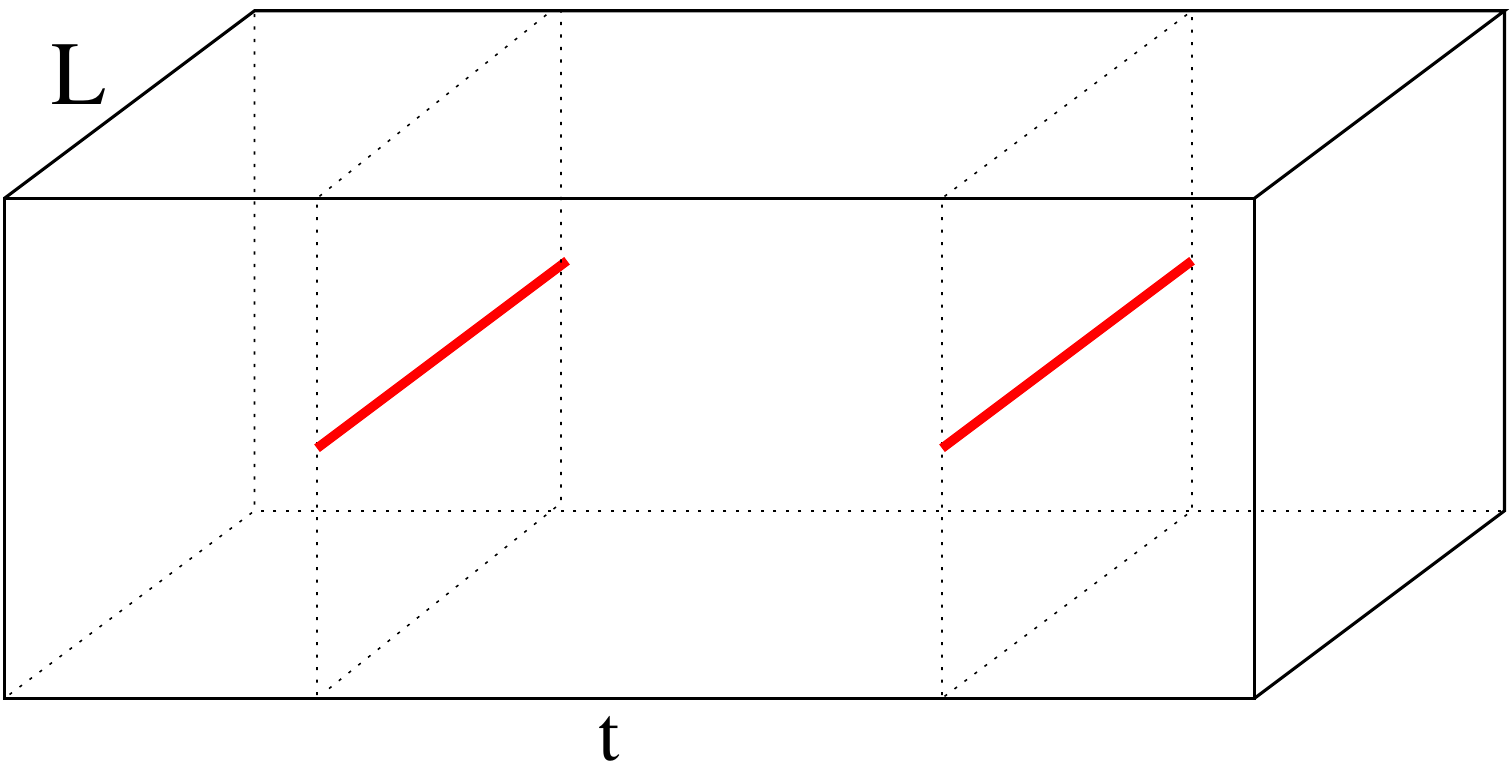}
}
\caption{Torelon correlator}
\label{torelon_diagram}
\end{figure}

The ``torelon'' is a gluonic excitation which is topologically non-trivial:
it is excited by any Wilson loop which wraps around the spatial boundary
in one direction, for instance, as illustrated Fig.~\ref{torelon_diagram},
\begin{align}
T_i(t) = \Tr \prod_{k=0}^{L-1} U_i(\vec{x}+k\hat{i},t),
\end{align}
where $i=1,2,3$ is one of the spatial directions and $\hat{i}$ the unit
vector in that direction.\footnote{We initially adopted periodic boundary
conditions in space and anti-periodic in time for the fermion fields.
However, the dynamical fermions drive $T_i$ to negative values in this
case, with a $Z_2$ degeneracy between the two complex $Z_3$ sectors.
We occasionally observed tunneling of $T_i$ between these two sectors,
and long metastabilities. For this reason, we changed the spatial 
boundary conditions to anti-periodic, which makes $\langle T_i \rangle$
real positive.} We extract the mass of this excitation from 
the exponential decay of the correlator $\langle T_i(0)^* T_i(t) \rangle$.
To suppress excited states, we smear the links within each time-slice
before constructing $T_i$. This observable has been used for a long time
to extract the string tension $\sigma$ in Yang-Mills theories~\cite{Michael1989}:
it can be viewed as a loop of gluonic string, whose energy $m_T(L)$ grows with
its length as $\sigma L$. So our initial, naive expectation was to measure
a mass $m_T(L)$ growing with $L$, until perhaps the string would break due
to fermion-pair creation.

\begin{figure}
\includegraphics[width=0.49\textwidth]{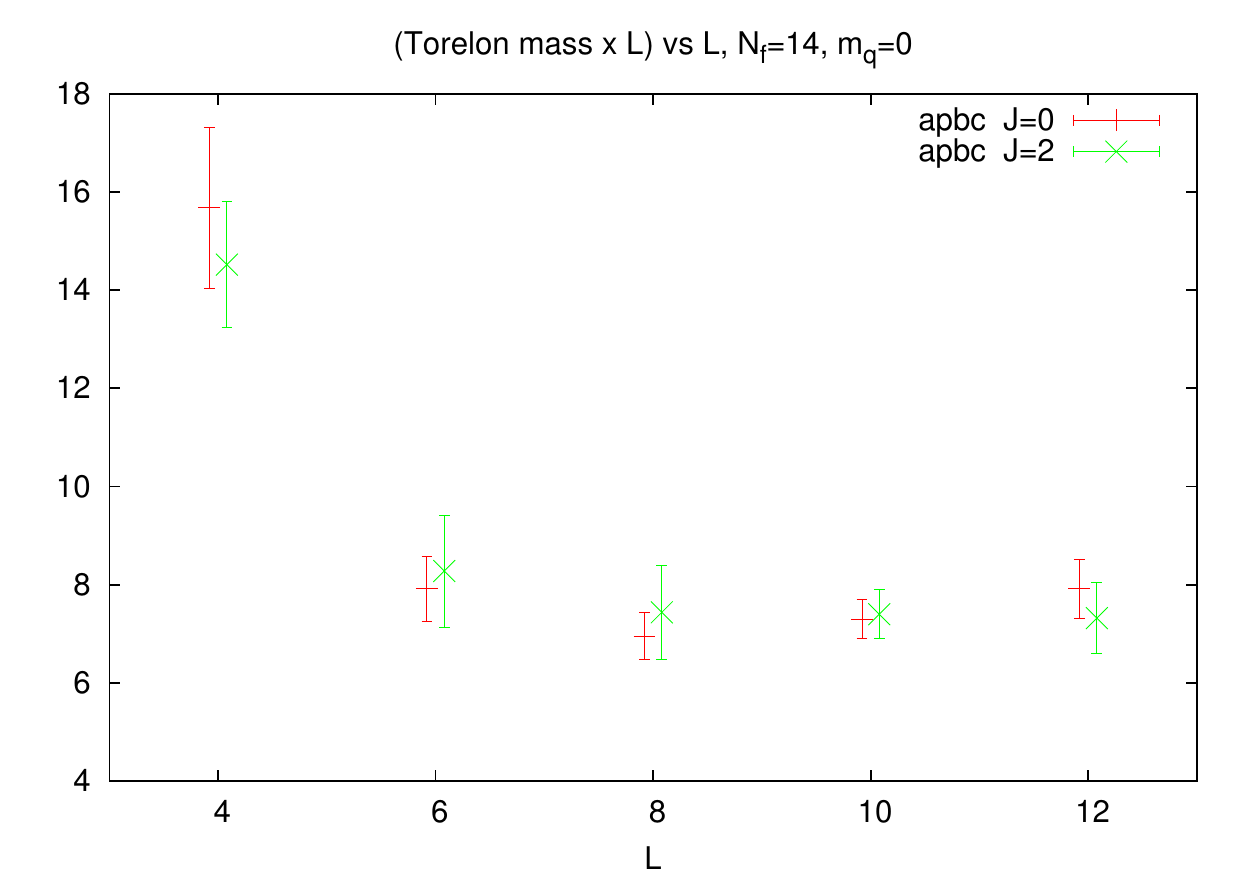}
\includegraphics[width=0.49\textwidth]{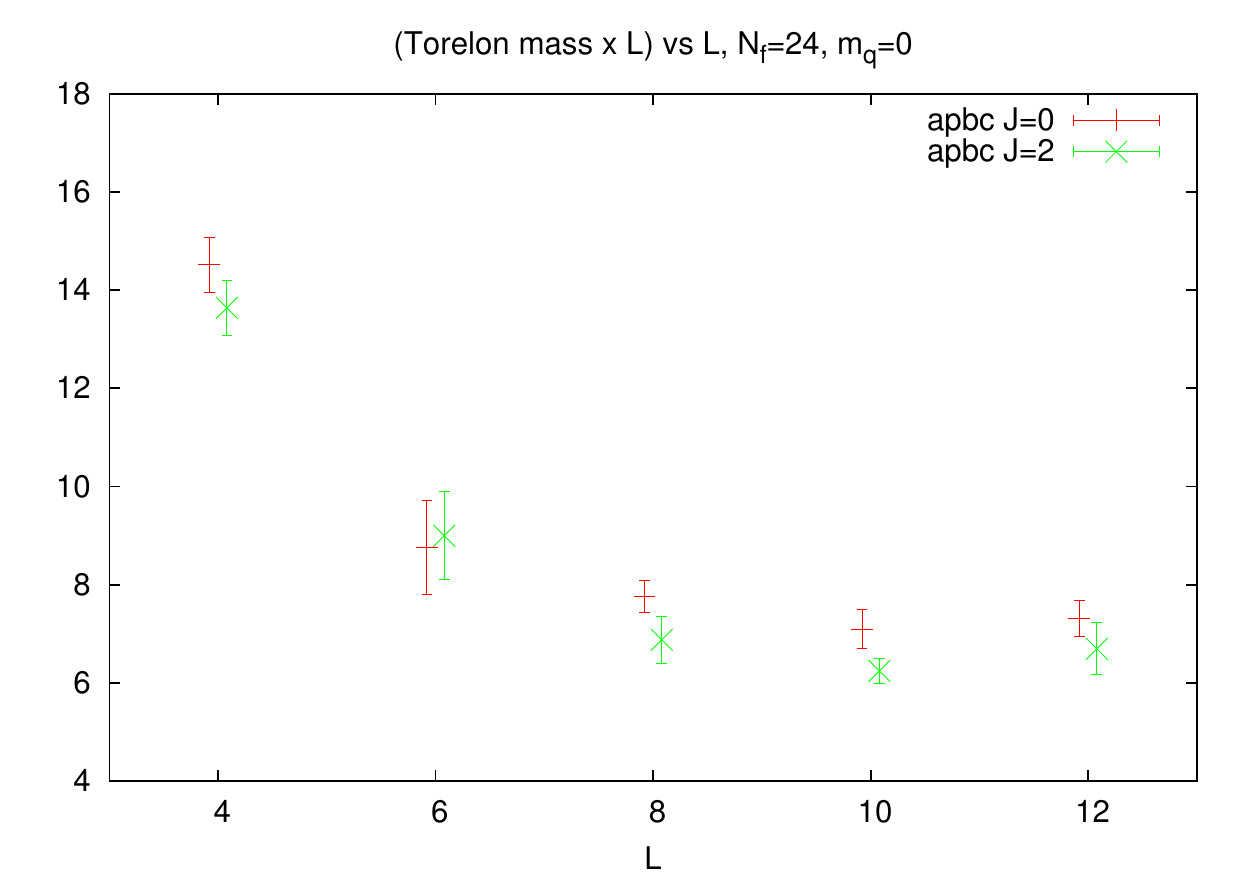}
\caption{The torelon mass $m_T(L)$ multiplied by $L$ versus $L$, left: $\hatNf=14$, right: $\hatNf=24$ (for anti-periodic boundary conditions).}
\label{torelon}
\end{figure}

This is not at all what we observed. The dimensionless quantity which we
measure, $a m_T(L)$, {\em decreases} on larger lattices corresponding
to a larger ratio $L/a$. Clearly, our theory is not confining. 
Moreover, as shown Fig.~\ref{torelon}, the combination $L m_T(L)$ is
approximately constant as $L$ is increased. So the torelon mass varies
as $1/L$ (actually, for small $L$ it initially decreases even faster with $L$
as seen in the figure). Thus, there is no intrinsic mass scale which appears
in this channel: the torelon mass is set by the system size $L$.
This remarkable result is our first evidence that our theory is IR-conformal.

We actually combined the $T_i(t), i\in\{1,2,3\}$ into two representations of the
cubic discrete rotation group: the $A_1^+$ (corresponding to a $0^+$ 
representation of $O(3)$) and the $E^+$ (corresponding to a $2^+$ representation
of $O(3)$). The mass of the $E^+$ seems to be slightly smaller, as
observed in small-volume analytic Yang-Mills calculations~\cite{Luscher}.

\begin{figure}
\centerline{
\includegraphics[width=0.49\textwidth]{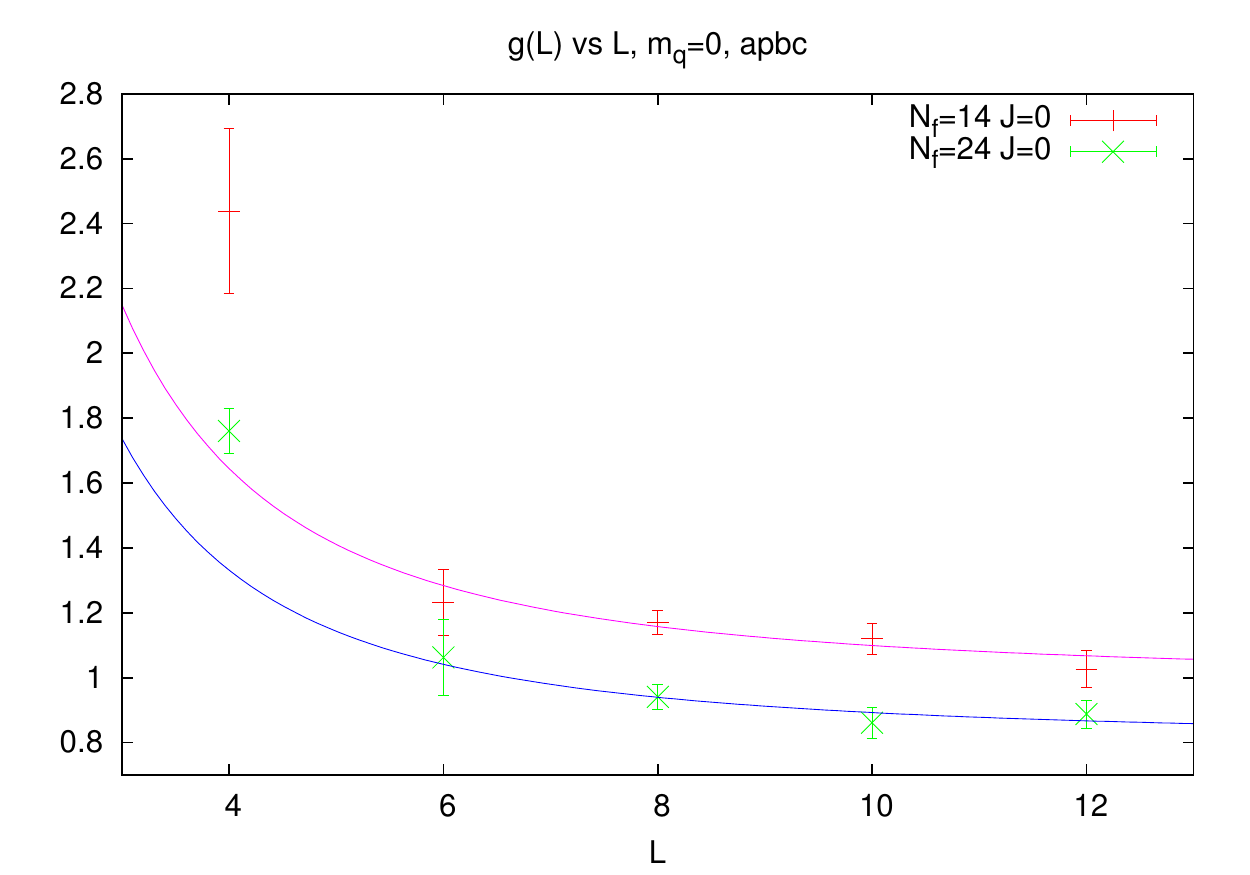}
}
\caption{The running coupling $g(L)$ defined via the temperature-dependence of 
the Debye mass, which is identified with the torelon mass, for $\hatNf=14$ and 24.
For each $\hatNf$, the curve is a fit to a constant plus $(a/L)^2$ corrections of 
the 4 largest volumes. The larger $\hatNf$ has a smaller coupling.}
\label{running_g}
\end{figure}

Now, by relabelling the spatial direction $i$ as the imaginary time direction,
one realizes that we are measuring the correlation of two time-like Polyakov
loops, whose decay rate is governed by the Debye mass, given perturbatively 
at lowest order by

\begin{align}
m_D(T) = 2 g T \sqrt{\frac{\Nc}{3} + \frac{\Nf}{6}}
\end{align}

This expression allows us to {\em define} a running coupling $g(L)$ via

\begin{align}
g(L) \equiv \frac{m_T(L) L}{2 \sqrt{\frac{\Nc}{3} + \frac{\Nf}{6}}}
\end{align}

and we see that, in this scheme, our running coupling seems to go to a
non-zero constant as $L$ increases (although one cannot exclude, of course,
that it slowly goes to zero). Therefore, we have numerical evidence
supporting the view
that our strong-coupling, chirally symmetric theory is IR-conformal and
non-trivial.

Interestingly, the extracted value of $g(L)$ approaches $\sim 0.95$ and 
$\sim 0.80$ for $\hatNf=14$ and $24$ respectively. So the IR fixed-point coupling
value decreases as $\hatNf$ increases. This is what one would expect: as $\hatNf$ 
keeps increasing, the ordering effect of the fermions increases, and all
Wilson loops are driven towards 1, their free field value. At the same time,
any definition of a running coupling will approach zero. The theory 
becomes trivial for $\hatNf\to\infty$, even in the strong-coupling limit.
We will come back to this point in Sec.~IV.

\begin{figure}[htbp!]
\includegraphics[width=0.49\textwidth]{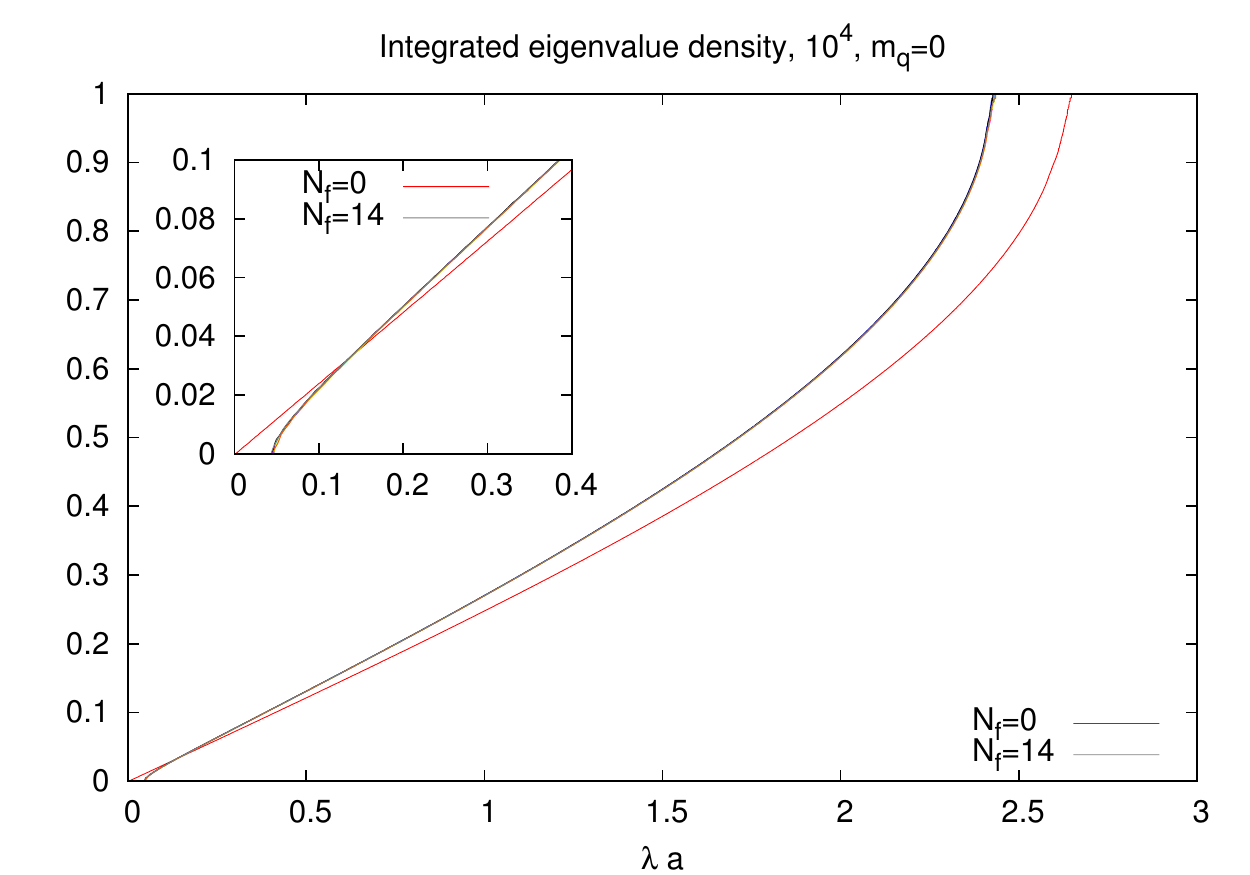}
\includegraphics[width=0.49\textwidth]{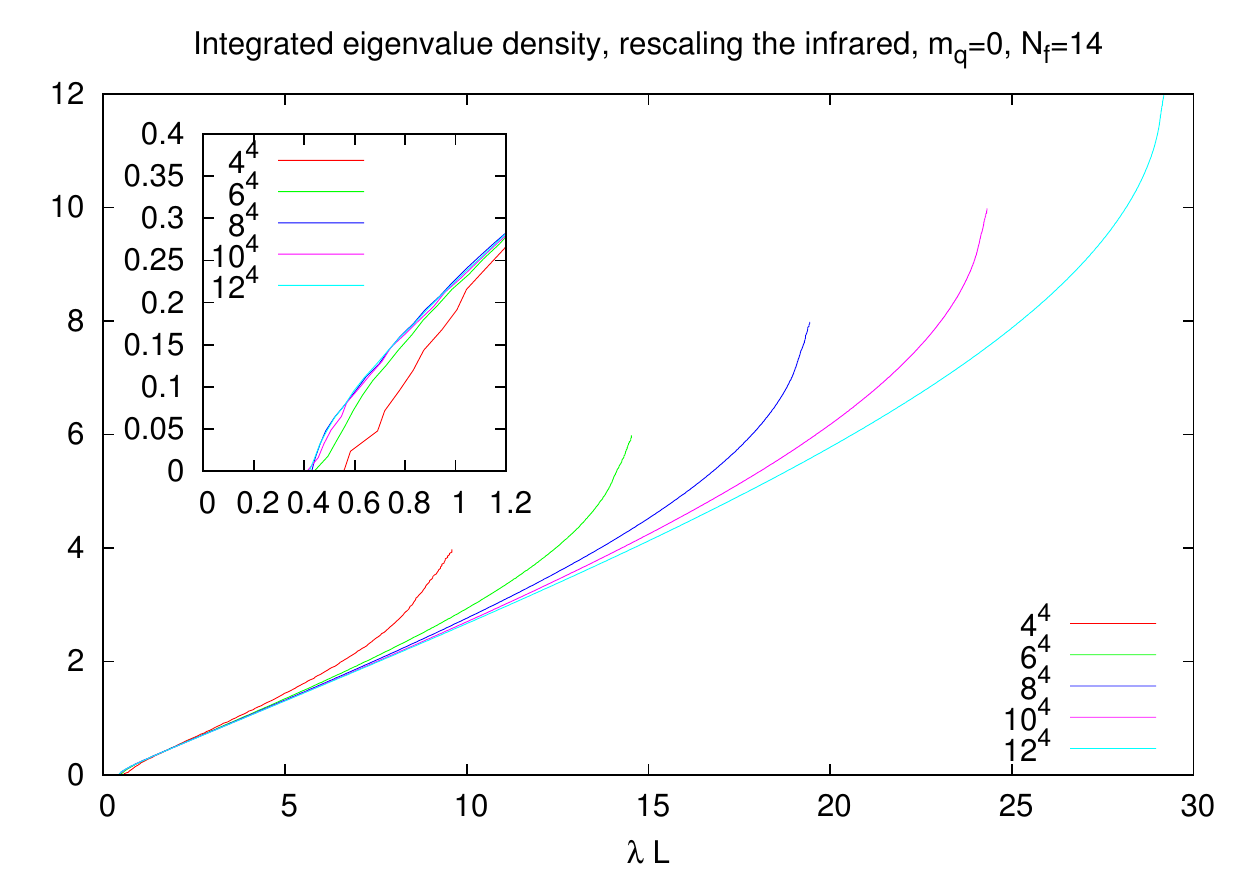}
\caption{The integrated eigenvalue density. Left: comparison of $\Nf=0$ (quenched) with $\hatNf=14$ in the chirally restored phase, where only the latter shows a spectral gap.
Right: The rescaled spectral gap, indicating $1/L$ scaling.}
\label{dirac}
\centerline{
\includegraphics[width=0.49\textwidth]{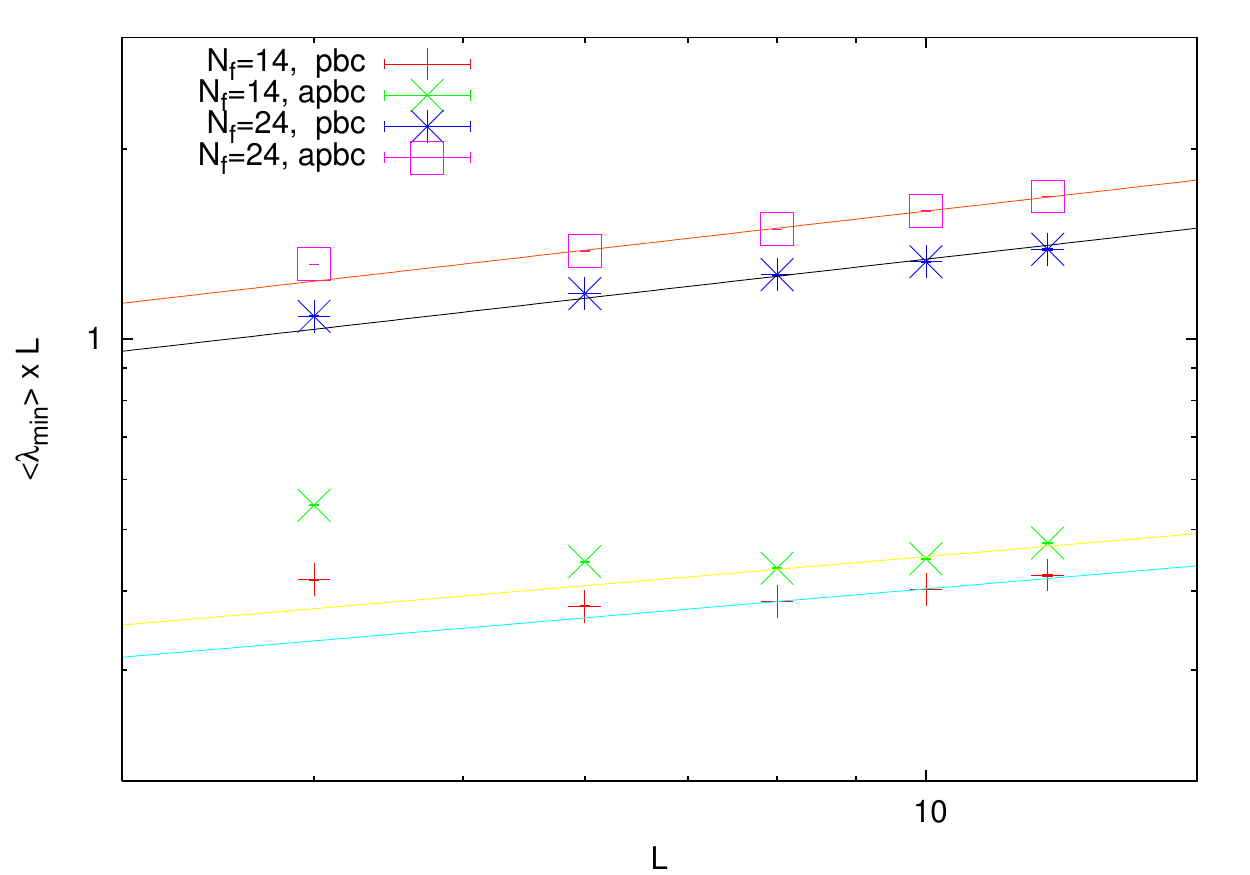}
}
\caption{The anomalous dimension from fits to the spectral gap for $\hatNf=14, 24$ with periodic and anti-periodic boundary conditions. Fitted exponents to the
three largest volumes are 
$\gamma^*\sim 0.26$ for $\hatNf=14$ and $\gamma^*\sim 0.38$ for $\hatNf=24$.
}
\label{gapNf}
\end{figure}

\subsection{Characterizing the chirally restored phase: (II) Dirac Spectrum}

We now turn to fermionic properties, starting with the spectrum of the
Dirac operator.
We have analyzed the Dirac eigenvalue spectrum of the configurations in our
Monte Carlo ensembles, using a Lanczos algorithm to obtain an approximation
of the whole spectral density and an Arnoldi method to extract the smallest
eigenvalues to high accuracy.
The observable shown in Fig.~\ref{dirac} is the integrated eigenvalue density, 
defined as:
\begin{align}
\int_0^\lambda \rho(\bar{\lambda})d\bar{\lambda}=\frac{\rank(\lambda)}{\rank(\text{Dirac matrix})}  \in [0,1].
\end{align}
This function of $\lambda$ counts the fraction of eigenvalues smaller than 
$\lambda$. Its derivative is simply the eigenvalue density $\rho(\lambda)$.
We first compare this observable on quenched configurations ($\Nf=0$) 
and in the chirally symmetric phase ($\hatNf=14$).
In Fig.~\ref{dirac} (left), 10 curves for 10 configurations are superimposed:
variations in the spectrum are very small.
We observe that $\Nf=0$ and $\hatNf=14$ spectra are similar in the ultraviolet, 
but differ in the infrared, as illustrated in the inset. 
The $\Nf=0$ curve starts linearly from the origin, 
reflecting an eigenvalue density approximately constant near $\lambda=0$.
On the contrary,
the integrated eigenvalue density for $\hatNf=14$ shows a spectral gap for small eigenvalues, 
which is of course consistent with chiral symmetry restoration according to the Banks-Casher relation, since $\rho(0)=0$.

The crucial question is on which scale does this spectral gap depend. 
To answer this question, in Fig.~\ref{dirac} (right) we compare the integrated 
eigenvalue density at $\hatNf=14$ for various lattice volumes, 
$L=4,6,8,10$ and $12$. As evidenced in the inset, we find that the spectral gap scales $\propto 1/L$ to a good approximation, 
which is a strong indication that our theory is IR-conformal\footnote{
If one would take the limits $L\to\infty$ first, then $m_q\to 0$, the expected
spectral density for a conformal theory would be $\rho(\lambda) \sim
\lambda^{(3-\gamma^*)/(1+\gamma^*)}$. Here, we take the opposite order of
limits.}: 
There does not seem to be any length scale in the chirally restored phase other than the box size $L$.

Actually, small deviations from $1/L$ scaling allow us, in principle, to 
extract the anomalous mass dimension $\gamma^*$. We make such an attempt
in Fig.~\ref{gapNf}, where the gap has been multiplied by $L$ already:
deviations from a constant are indicative of anomalous dimension, {\em provided}
other corrections $\Ord((a/L)^2)$ are negligible. 
The effect of a finite system size $L$ on the Dirac spectrum has not been 
analyzed yet. We have simply considered that the infrared conformal symmetry 
is explicitly broken by the infrared scale $1/L$, which is the analogue of
an explicit breaking by a quark mass $m_q$. Consequently, we expect the
mass gap to behave as $(1/L)^{1/(1+\gamma^*)}$.
A crude, 2-parameter fit based on our 3 largest volumes gives $\gamma^* \sim 0.26$ and $0.38$
for $\hatNf=14$ and $24$, respectively.
Simulations on larger volumes should be performed to bring under control 
the systematic error in these estimates. The true, infinite volume value of
$\gamma^*$ seems to be approached from below.

\subsection{Characterizing the chirally restored phase: (III) Hadron Masses}

\begin{figure}
\includegraphics[width=0.49\textwidth]{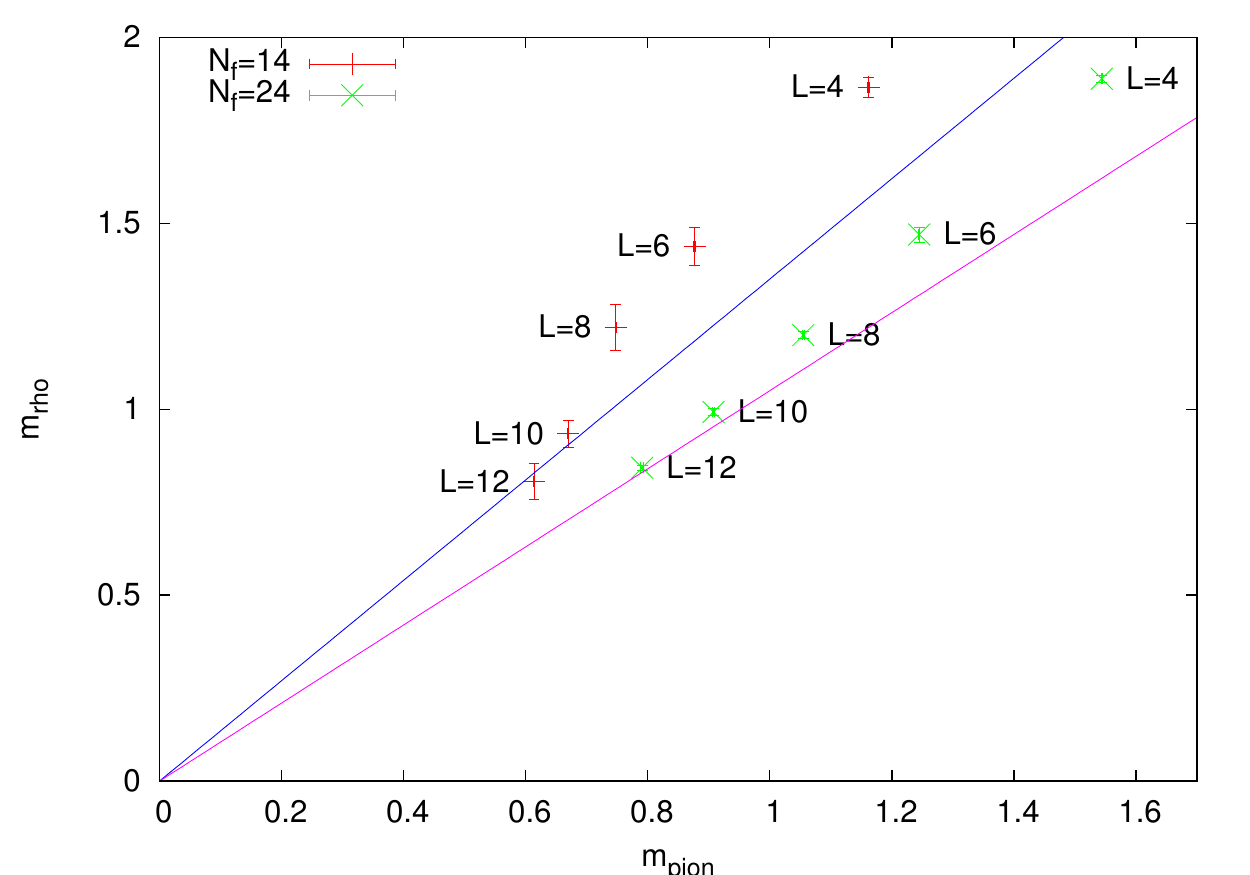}
\includegraphics[width=0.49\textwidth]{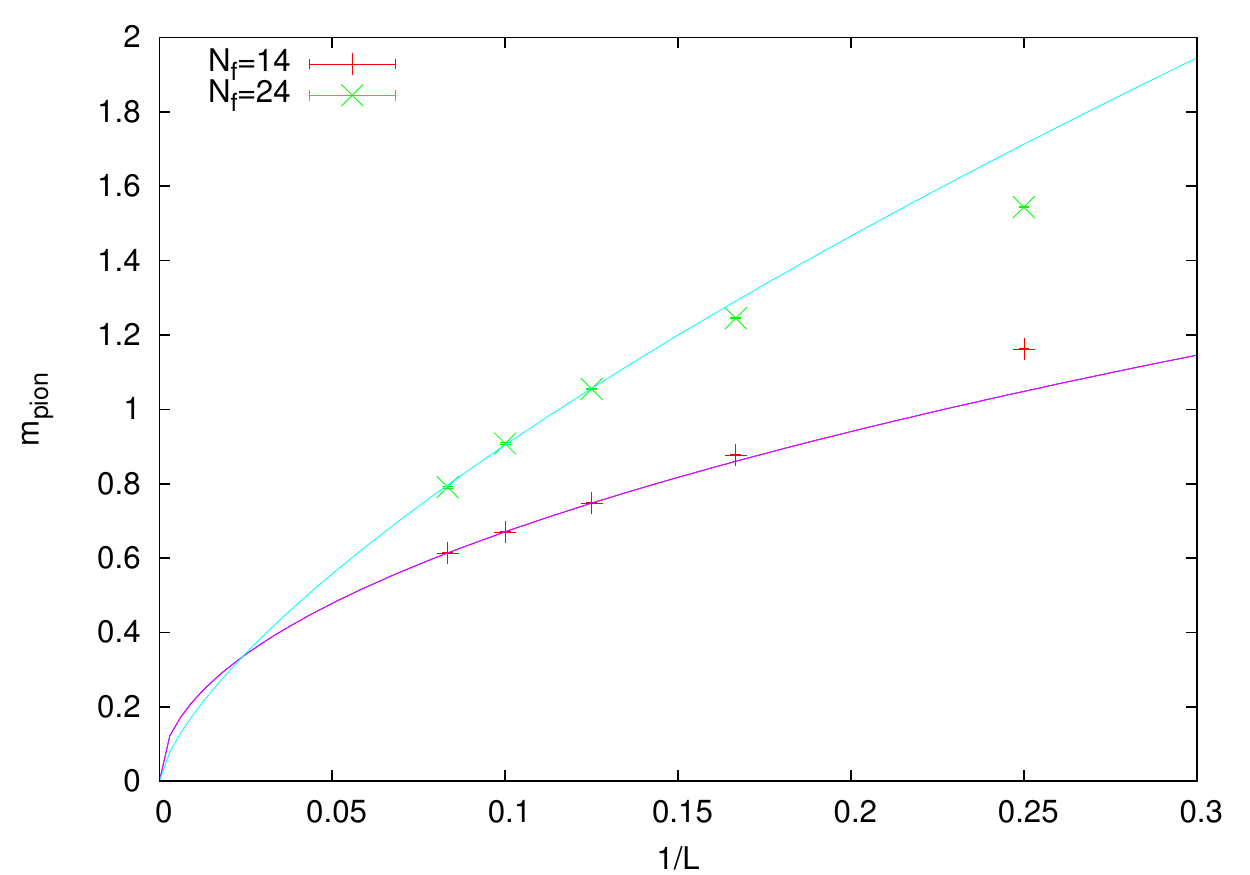}
\caption{Left: meson masses $m_\pi$ and $m_\rho$ for various system sizes $L$,
showing a fast decrease of the masses as $L$ increases, while the mass ratio
remains more or less constant. Right: pion mass as a function of $1/L$,
with the three largest volumes fitted by the ansatz $(1/L)^{1/(1+\gamma^*)}$,
yielding $\gamma^* \sim 1.0$ and $0.4$ for $\hatNf=14$ and $24$, respectively.
}
\label{hadron}
\end{figure}

Finally, we turn to hadron masses measured on our $\hatNf=14$ and $\hatNf=24$ Monte Carlo ensembles.
Even though the quark mass is zero in these ensembles, we observe non-zero
hadron masses. As expected, parity partners are degenerate since chiral 
symmetry is restored.
Now, if our theory is IR-conformal, the masses which we measure are exclusively
due to finite-size effects: all masses should go to zero as the lattice size
$L$ is increased. This is what we observe, as shown in Fig.~\ref{hadron} (left):
the ``pion'' and ``rho'' masses both decrease by a factor $\sim 2$ as the
lattice size is increased from $L=4$ to $12$. Notice however that the smallest
mass is still $> 0.6$, which is not very light.

Furthermore, one generally expects that the approach to zero should be the same
for all hadrons, so that mass ratios should remain constant as $L\to\infty$.
Note that there may be exceptions to this ``rule'': 
in the $2d$ $O(3)$ sigma model near $\theta=\pi$, the mass of the $O(3)$ singlet
state approaches zero faster than that of the $O(3)$ triplet as $\theta$
approaches $\pi$~\cite{Controzzi}.
Here, our Fig.~\ref{hadron} (left) would show all data points aligned on ``rays'' going
through the origin if mass ratios were constant. One can see deviations from
this behavior, which perhaps are caused by the not-too-light masses which 
we measure. Another possible cause is technical: as $L$ is increased, the
groundstate masses in each channel decrease, but so do also the mass differences
between groundstate and excited states. It becomes more difficult to extract
the groundstate mass, and our lattices are likely too short to properly
control this source of systematic error.

Nevertheless, we show in Fig.~\ref{hadron} (right) the mass of the ``pion'', in
which we have the most confidence, as a function of $1/L$.
Since $1/L$ breaks the conformal symmetry much like a quark mass $m_q$ would,
we expect that the pion mass should scale the same way, namely as 
$(1/L)^{1/(1+\gamma^*)}$, if $L$ is large enough. A 2-parameter fit
to our three largest system sizes gives $\gamma^* \sim 1.0$ and $0.4$
for $\hatNf=14$ and $24$, respectively. 
As with the estimates of $\gamma^*$ from the Dirac spectrum, simulations on
larger volumes are needed to bring the systematic error under control.

\section{The conjectured phase diagram}

\begin{figure}
\includegraphics[width=0.49\textwidth]{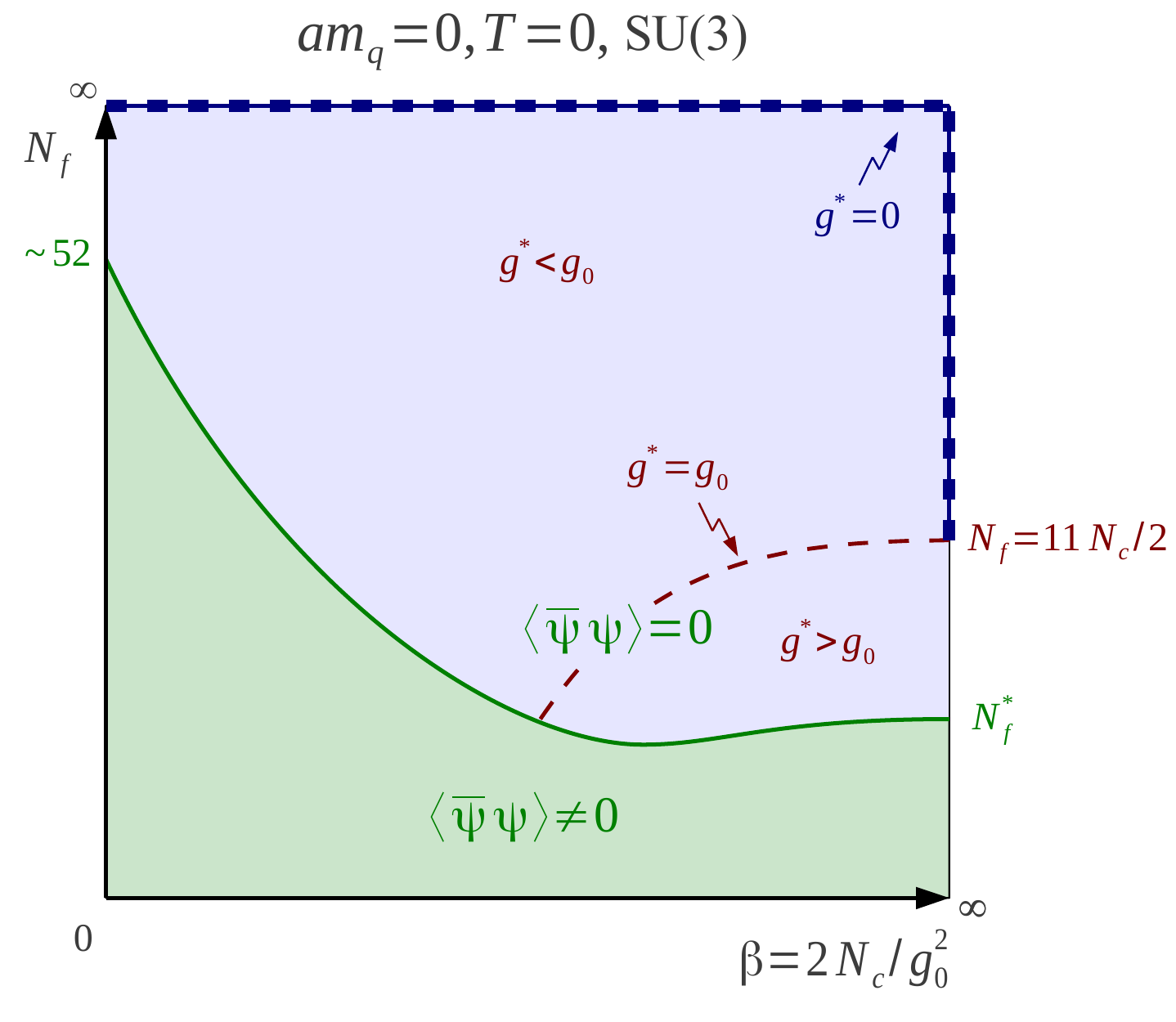}
\includegraphics[width=0.49\textwidth]{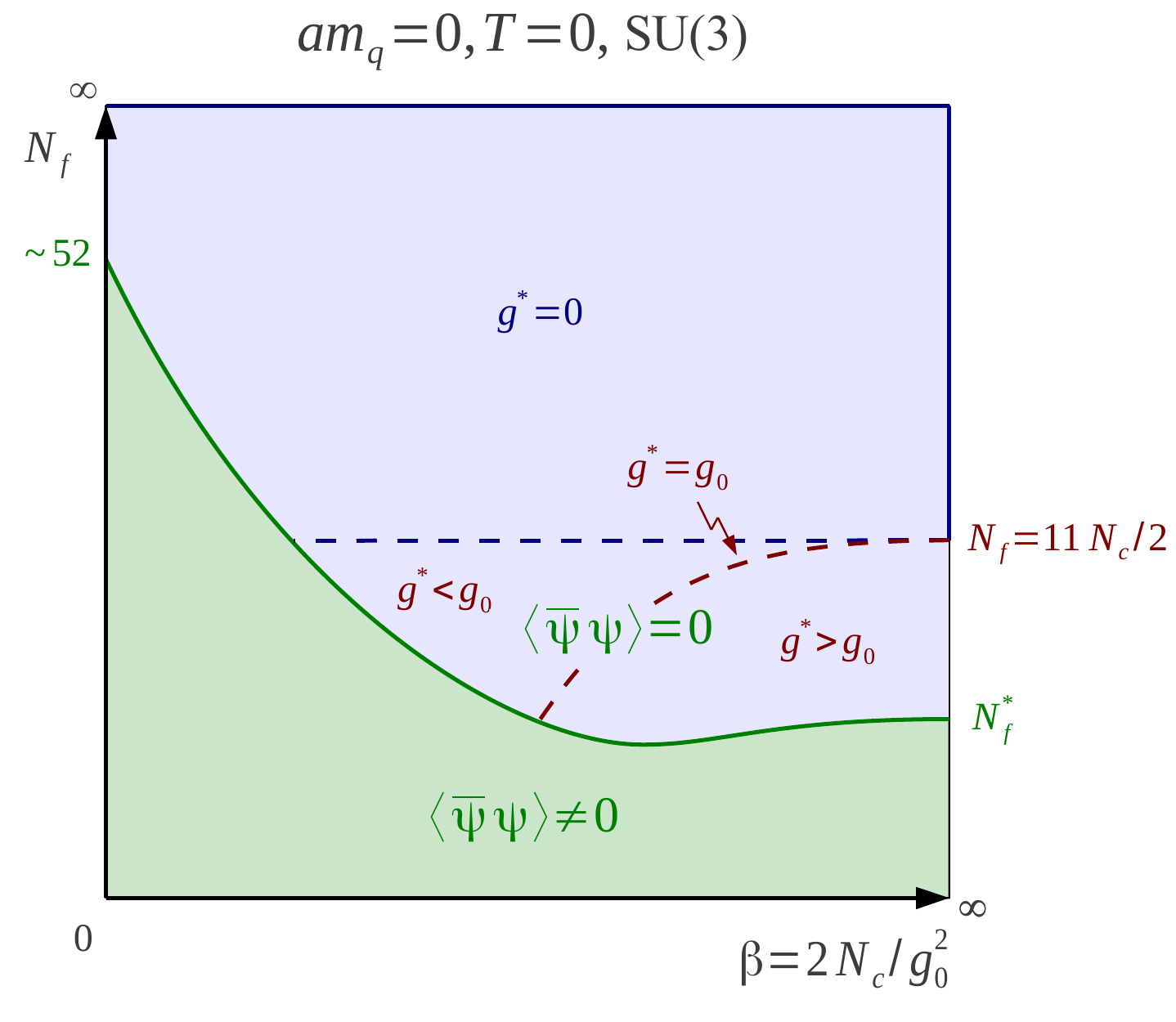}
\caption{Left: conjectured phase diagram in the $(\beta,\Nf)$ plane. A single
phase transition line separates the chirally broken phase from the chirally
symmetric, IR-conformal phase. The thick dotted line indicates trivial theories.
Right: alternative scenario, where the trivial theories extend to the area above $\Nf=11\Nc /2$. It is not favored by our measurements.}
\label{conjectured}
\end{figure}

Using both gluonic and fermionic observables,
we have presented evidence that the chirally restored phase at strong coupling 
is IR-conformal and non-trivial, and speculated about a connection to the 
conformal window in the continuum.
We now want to propose a phase diagram sketched in Fig.~\ref{conjectured} (left),
as a function of the plaquette coupling $\beta=6/g_0^2$ and of the number of
would-be fundamental flavors $\Nf$ in the weak-coupling limit $\beta=\infty$.
That is, we simply convert the number $\hatNf$ of staggered fields to 
$\Nf=4\hatNf$. Moreover, we promote $\Nf$ to a real, continuous parameter:
while $\Nf$ must be integer for a well-defined continuum theory, one may
let it take any value in the statistical model defined by the lattice 
partition function.
Our conjectured phase diagram can be compared with, e.g., those of 
Ref.~\cite{Banks-Zaks,Miransky-Yamawaki}: one can see substantial differences.
The essential feature of our phase diagram is that the $\beta=0$ IR-conformal 
phase is analytically connected with the weak-coupling, continuum IR-conformal 
phase. This is the simplest scenario, supported by our exploratory scan
in $\beta$ shown Fig.~\ref{fixedam}. A single phase transition line 
$\Nf^c(\beta)$ separates the region of broken chiral symmetry at small $\Nf$ 
from the chirally symmetric region at large $\Nf$. The transition is 
first-order, at least for some range of $\beta$ starting from zero.
Moreover, the number $\hatNf^c$ of staggered fields which bring enough order
to restore chiral symmetry at $\beta=0$, $\hatNf^c(\beta=0) = 
1/4 \Nf^c(\beta=0) = 13(1)$, is remarkably close to the expected number 
$\Nf^*$ of continuum quark fields which achieve the same effect\footnote{Different
groups place $N_f=12$ above or below $\Nf^*$.}.
This may be more than a numerical accident. At strong coupling, taste symmetry
breaking is maximal, and $\hatNf=13$ staggered fields can be viewed as $\hatNf$
massless fields, plus $3\hatNf$ fields with mass $\Ord(a^{-1})$.
Only the former have a significant effect toward chiral symmetry restoration.

In the chirally symmetric phase, we see no evidence for a dynamically
generated mass scale of any sort.
Then, based on our results for the running coupling Fig.~\ref{running_g},
we conjecture that for any finite $\beta$ and $\Nf>\Nf^c(\beta)$, 
large-$\Nf$ lattice QCD is IR-conformal, with a non-trivial fixed-point 
coupling $g^*>0$. This value changes continuously with $(\beta,\Nf)$, reaching 
the value zero for $\beta=\infty, \Nf > 33/2$ and for $\Nf=\infty~\forall 
\beta$,
as indicated Fig.~\ref{conjectured} (left) by a thick dotted line. $g^*$ grows
as one moves away from this dotted boundary towards the phase transition line.

An alternative scenario would be that the running coupling in Fig.~\ref{running_g}
slowly approaches zero, and $g^*=0$ for $\Nf > 33/2$ for any
$\beta$ in the chirally symmetric phase. This is sketched in Fig.~\ref{conjectured} (right). If the basin of attraction of the
weak coupling trivial fixed point would extend all the way to the strong
coupling limit, one should observe for the running coupling
$g^2(L) \sim 1/\log(L/L_0)$, with $L_0 \sim \Ord(a)$.
Whether or not this happens depends on the marginal operators induced by our
lattice action. 
Our numerical results Fig.~\ref{running_g} for the running coupling
are indeed consistent with this possibility. But we should also observe
in this case an anomalous mass dimension 
$\gamma^*=0$, which is not favored by our other measurements.
More careful, large-scale simulations are necessary to settle this issue.

Finally, one may consider the line $g^*=g_0$, with $g_0=(2\Nc/\beta)^{1/2}$, 
where the IR fixed-point coupling has the same value as the bare coupling, 
so that the coupling does not run as a function of the renormalization scale.
This line starts at the point $(\beta=\infty,\Nf=33/2)$ where $g^*=g_0=0$.
Its precise location depends on the chosen renormalization scheme.
It is not associated with any kind of singularity of the free energy.
There is no phase transition along this line: 
simply, on the left (resp. on the right) of that line, the coupling increases
(resp. decreases) from $g^*$ as one reduces the distance scale. Since there
is a lower distance cutoff $a$ no divergence is observed as one crosses this
[scheme-dependent] line.

\begin{figure}
\centerline{weak coupling:\hspace{7cm}strong coupling:}
\includegraphics[width=0.49\textwidth]{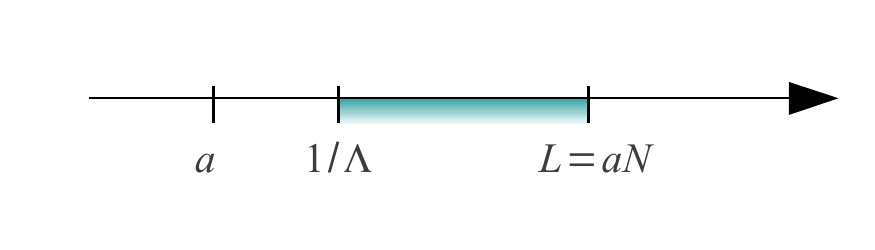}
\includegraphics[width=0.49\textwidth]{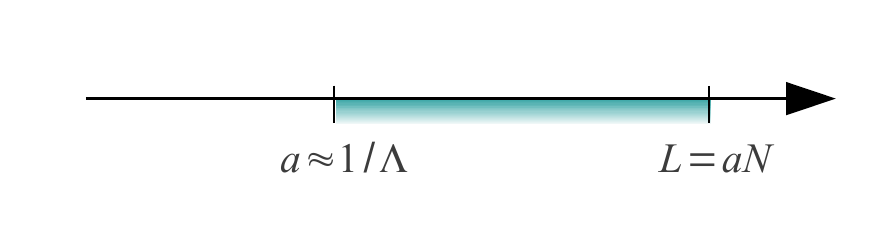}
\caption{
The ordering of scales at weak coupling (left) and strong coupling (right), showing that the range of conformal invariance is larger in the latter case.}
\label{scales}
\end{figure}

We have determined the phase diagram in the strong-coupling region only.
Studying the continuum limit is of course much more difficult, due to the 
large lattices that have to be used in order to control the finite size 
effects and the difficult control of lattice artifacts. 
We would like to suggest that the strong-coupling limit may represent
an advantageous ``poor man's laboratory'' for the study of $4d$ IR-conformal
gauge theories.
In particular, as illustrated Fig.~\ref{scales}, the range of 
scales over which conformal invariance applies
for a given computing effort is greatly reduced at weak coupling: there, 
for a given lattice size $N^4$, the scales are ordered as follows:
\begin{align}
a \ll 1/\Lambda \ll L=Na.
\end{align}
where $\Lambda$ is the scale generated by the asymptotically free gauge 
dynamics and Yang-Mills perturbation theory applies at distances $\lesssim 1/\Lambda$. In contrast, at strong coupling, where the lattice becomes 
maximally coarse, there is no small distance where Yang-Mills perturbation theory applies, and the hierarchy is:
\begin{align}
a \sim 1/\Lambda \ll L=Na.
\end{align}
Hence the dynamical range of conformal invariance, characterized by the product 
$L\Lambda$, is maximized at $\beta=0$ for a given lattice size $N=L/a$.

\section{Conclusion}

We have shown that for $\beta=0$, contrary to common wisdom, there exists a 
strongly first-order, $\Nf$-driven bulk transition to a chirally symmetric phase. 
In the chiral limit, the transition occurs for $\hatNf^c=13(1)$ staggered fields, i.e. $\Nf^c=52(4)$ \emph{continuum} flavors.
This finding is in contrast to the mean-field prediction, whose failure can
be traced back to approximations relying on $\Nf$ being small.
Clearly, the conventional, automatic association of the strong-coupling limit 
with confinement and chiral symmetry breaking is too naive.
Furthermore, the chirally restored phase extends to weak coupling.

We have also shown numerical evidence that the $\beta=0$ chirally restored phase of ``large-$\Nf$ QCD'' is IR-conformal, with a non-trivial, $\Nf$-dependent value of the IR fixed-point coupling.
We conclude that the strong-coupling limit is the laboratory of choice to study a 4d IR-conformal gauge theory.
Simulations at large $\Nf$ and {\em zero} quark mass can be performed without 
too much computational effort since a gap appears in the Dirac spectrum.
As $\Nf$ increases, the spectral gap increases, the average plaquette
approaches 1, and the fixed-point coupling approaches 0.
Setting the quark mass to zero eliminates one IR scale, leaving the system
size $L$ as the only remaining one. This greatly simplifies the analysis
of simulation results.

Since we have not observed any evidence for an additional $T=0$ phase transition as $\beta$ is increased, we speculate that the strong coupling chirally symmetric, IR-conformal phase is
analytically connected with the continuum IR-conformal phase.\\

One may ask how robust these statements are with respect to the particular
discretization of the Dirac operator and the gauge action. While the 
quantitative details of the phase transition $\Nf^c(\beta)$ will surely change,
we think that the qualitative features will remain. Chiral symmetry breaking
at strong coupling, for small $\Nf$, is a general consequence of the disorder
in the gauge field. The ordering effect of fermions also is generic. So we
do expect a bulk transition, generically of first-order, as a function of 
$\Nf$ in the strong-coupling limit. Actually, such a transition was observed
for Wilson fermions in Ref.~\cite{Iwasaki, Nagai}. At intermediate coupling, additional
transitions may appear depending on the lattice action.
Interestingly, a first-order transition to a chirally broken phase as $\beta$ 
is reduced has been observed many times, for various lattice 
actions~\cite{Kogut1987,Damgaard1997,Deuzeman2010,Jin2011,Hasenfratz2012}.
These transitions were observed for some fixed value of $\Nf$. Here, we simply
put all these earlier observations together.
It is interesting that this phase transition is consistently of first-order.
If the first-order nature persists all the way to the continuum limit,
then walking dynamics will not be observed, and the transition to the
conformal window will be characterized by ``jumping dynamics'', as proposed
by Sannino~\cite{Sannino2012}.

One may also wonder what happens to the $(\beta,\Nf)$ phase diagram as the gauge
group or the fermion content is changed. For $4d$ compact $U(1)$, the change
would not be dramatic, because the strong-coupling behavior is much the
same as for $SU(3)$: our 
first-order transition line would end at $(\beta\approx 1.01,\Nf=0)$ on
the horizontal axis rather than on the vertical axis, and the region of 
triviality would cover the whole chirally symmetric phase, except for $\Nf=0$.
For $SU(2)$ or $SU(3)$ with adjoint fermions, the change would be more
significant: in the strong-coupling limit, increasing $\Nf$ would order
the plaquette in the adjoint representation, not in the fundamental.
Center monopoles would likely condense~\cite{SO3}, and might delay or 
prevent the restoration of chiral symmetry.

\begin{figure}
\includegraphics[width=0.49\textwidth]{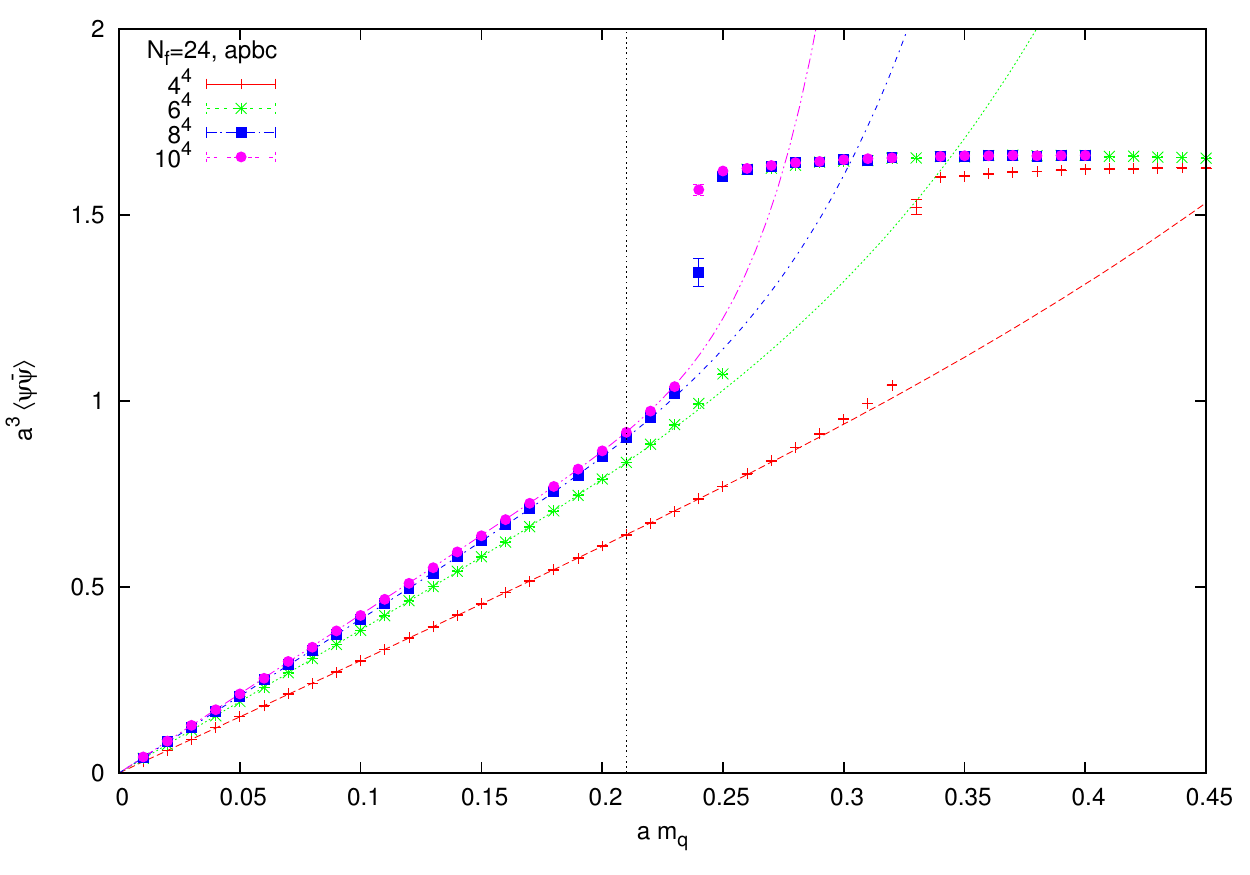}
\includegraphics[width=0.49\textwidth]{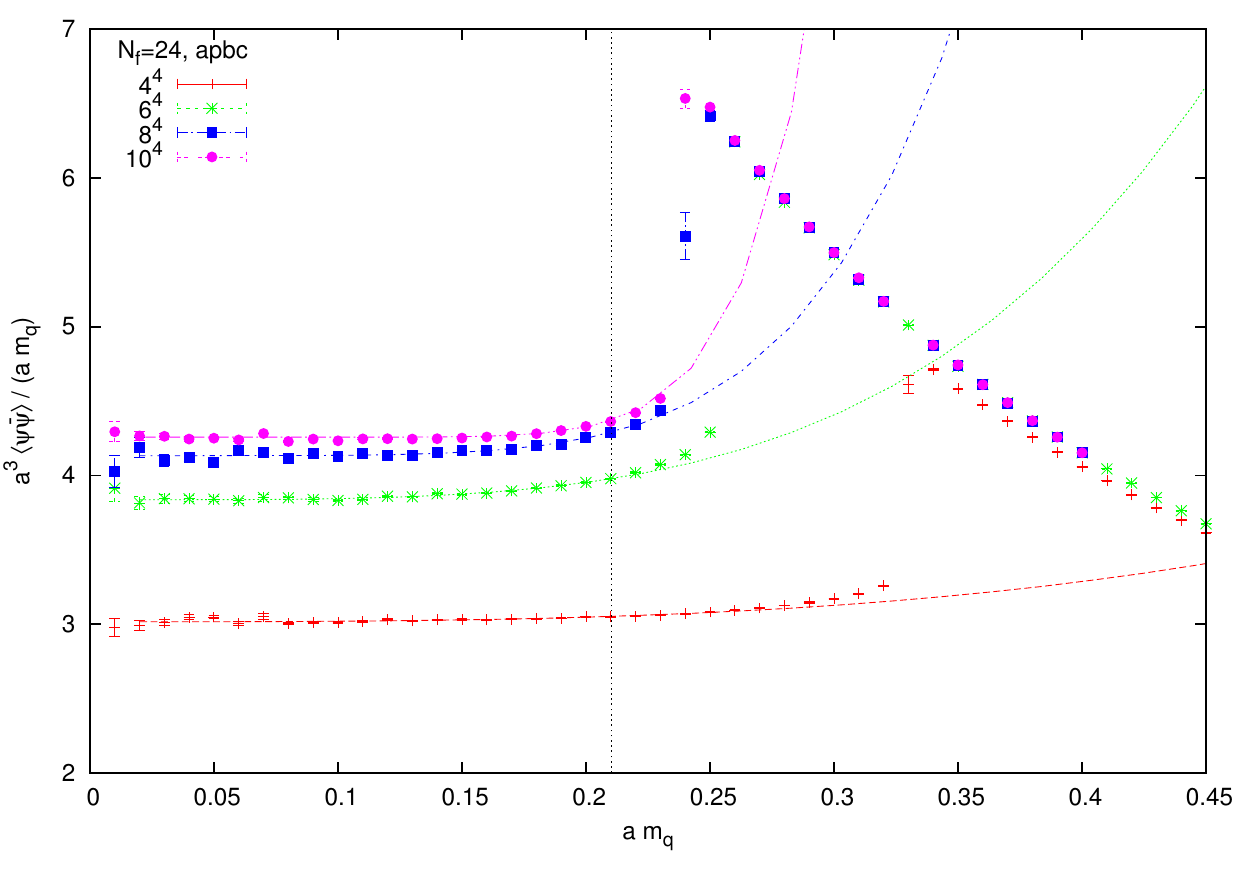}
\caption{Left: The chiral condensate as a function of the quark mass, showing clearly 
a linear relation as appropriate for the chirally symmetric phase. Deviations 
from linearity can be used to determine the mass anomalous dimension 
according to Eq.~\ref{ansatzgamma}.
Right: the same data, now divided by $(a m_q)$ to emphasize deviations from
linear behaviour. The fitted values of $\gamma^*$ depend on the fitting range
(delimited by the vertical line),
but tend to be negative (e.g. $\gamma^* \sim -0.5$ for $L=10$).}
\label{quarkmass}
\end{figure}

Finally, there are many directions in which to extend this exploratory study.
To buttress the claim that the chirally symmetric phase is IR conformal,
more observables, like the static potential and the Fredenhagen-Marcu order
parameter, should be studied. Also, and to make contact with other numerical
studies, a mass deformation could be introduced. As a first step in this
direction, we show in Fig.~\ref{quarkmass} the quark mass dependence of
the chiral condensate. This figure shows all the technical difficulties 
associated with extracting the anomalous dimension $\gamma^*$:
Heavier fermions have less ordering effect, which triggers a phase transition
back into the chirally broken phase for some critical fermion mass. 
Finite-size effects associated with that transition should not be 
included in the determination of $\gamma^*$. Moreover, the non-anomalous
contributions $\langle \bar\psi \psi \rangle = c_1 m_q + c_2 m_q^3$ easily
overwhelm the anomalous $m_q^{(3-\gamma^*)/(1+\gamma^*)}$. 
Actually, fits based on the ansatz
\begin{align}
 \langle \bar\psi \psi \rangle = c_1 m_q + c_2 m_q^{\frac{3-\gamma^*}{1+\gamma^*}} ~ ,
 \label{ansatzgamma}
\end{align}
or including an $m_q^3$ term~\cite{Smilga},
favor negative values for $\gamma^*$, which crucially depend
on the fitting range and the included analytic contributions.
We believe that extracting $\gamma^*$ from the Dirac spectrum provides better control
of the systematics, as emphasized in \cite{Patella}.
In any case, larger system sizes are required before reliable 
estimates of $\gamma^*$ can be obtained. This is beyond the scope of this work.
Here, we have argued that such reliable estimates can be obtained \emph{in principle} from the $L$-dependence of the
Dirac spectrum and of the meson spectrum, both measured at zero quark mass.

\section{Acknowledgements}

We are grateful for stimulating discussions with P.~Damgaard, L.~Del Debbio, A.~Hasenfratz,
A.~Kovner, A.~Patella, K.~Rummukainen, F.~Sannino, K.~Splittorff, B.~Svetitsky and R.~Zwicky.
Computations have been carried out on the Brutus cluster at the ETH Z\"urich and on a small cluster in the Sejong University physics department.
Ph.~de F.~thanks the Institute for Nuclear Theory at the University of Washington for its hospitality and the Department of Energy for partial support during the completion of this work.
S.~K.~is supported by Korea Foundation for International Cooperation of Science \& Technology (KICOS).
W.~U.~is supported by the Swiss National Science Foundation under grant 200020-122117.

\end{document}